\documentclass[3p,10pt,a4paper,twoside,fleqn,sort&compress]{filomat}
\usepackage{amssymb,amsmath,latexsym}
\usepackage[varg]{pxfonts}
\usepackage{subcaption} 
\newcommand{\FilVol}{xx}
\newcommand{\FilYear}{yyyy}
\newcommand{\FilPages}{zzz--zzz}
\newcommand{\FilDOI}{(will be added later)}
\newcommand{\FilReceived}{Received: dd Month yyyy; Accepted: dd Month yyyy}

\begin{document}

%

%
%
%

\author[affil1]{Oscar Romero}
\ead{oromero@dcom.upv.es}
\author[affil2]{Néstor Thome\corref{mycorrespondingauthor}}
\ead{njthome@mat.upv.es}



\title{Applications of Singular Entropy to Signals and Singular Smoothness to Images}


%
\address[affil1]{Departamento de Comunicaciones, Universitat Politècnica de  València, Camino de Vera s/n, Valencia, 46022, Spain}
\address[affil2]{Instituto Universitario de Matemática Multidisciplinar, Universitat Politècnica de  València, Camino de Vera s/n, Valencia, 46022, Spain}

\newcommand{\AuthorNames}{O. Romero, N. Thome}

\newcommand{\FilMSC}{Primary 15A18; Secondary 15A23.}
\newcommand{\FilKeywords}{Singular smoothness, information density, Singular Value Decomposition, images.}
\newcommand{\FilCommunicated}{Dijana Mosi\'c}
\cortext[mycorrespondingauthor]{* Corresponding author: Néstor Thome}
\newcommand{\FilSupport}{Research supported by Universidad Nacional de La Pampa (Grant Resol. 172/2024); by Grant PGI 24/ZL22, Departamento de Matemática, Universidad Nacional del Sur (UNS), Argentina; and by Ministerio de Ciencia, Innovación y Universidades of Spain (Grant Redes de Investigación, MICINN-RED2022-134176-T).}

\begin{abstract}
This paper explores signal and image analysis by using the Singular Value Decomposition (SVD) and its extension, the Generalized Singular Value Decomposition (GSVD). A key strength of SVD lies in its ability to separate information into orthogonal subspaces. While SVD is a well-established tool in ECG analysis, particularly for source separation, this work proposes a refined method for selecting a threshold to distinguish between maternal and fetal components more effectively. In the first part of the paper, the focus is on medical signal analysis, where the concepts of Energy Gap Variation (EGV) and Singular Energy are introduced to isolate fetal and maternal ECG signals, improving the known ones. Furthermore, the approach is significantly enhanced by the application of GSVD, which provides additional discriminative power for more accurate signal separation.
The second part introduces a novel technique called Singular Smoothness, developed for image analysis. This method incorporates Singular Entropy and the Frobenius norm to evaluate information density, and is applied to the detection of natural anomalies such as mountain fractures and burned forest regions. Numerical experiments are presented to demonstrate the effectiveness of the proposed approaches.
\end{abstract}

\maketitle

\makeatletter
\renewcommand\@makefnmark%
{\mbox{\textsuperscript{\normalfont\@thefnmark)}}}
\makeatother

\section{Introduction}\label{sec1}
When dealing with large volumes of data, the Fourier Transform \cite{CaMoVaSa,Co} and the Singular Value Decomposition (SVD) are two widely used techniques for obtaining low-order representations. These tools have found applications in diverse fields. For instance, in \cite{ElCoSt}, the SVD is studied in general, including how to compute it and discard singular values to reduce image size. In \cite{FrIn}, the authors use the SVD decomposition to decompose a Skew-symmetric finite game  into three subgames by means of vector space theory (symmetric, skew-symmetric, and asymmetric subspaces), to cite just a few examples of interesting applications. In the medical field, particularly in the analysis of electrocardiograms (ECG) from pregnant women, SVD has been commonly used to separate maternal and fetal signals \cite{CaMoVaSa}. It is also applied in tasks such as image deblurring \cite{HaNaOl} or grayscale image colorization \cite{SaMaAm}, where singular values help to determine optimal pixel matching from reference images. In tensor analysis, singular vectors are essential tools for decompositions beyond two dimensions \cite{RoSe}.
The GSVD extends the classical SVD and has shown utility in applications such as image encryption \cite{BaViGo} and ECG signal separation \cite{CaMoVaSa}. 

In \cite{ChLiHu}, an integrated method for fetal ECG extraction based on Adaptive Periodic Segment Matrix (APSM) and SVD is proposed, with promising results for clinical use. Similarly, SVD techniques are increasingly employed in EEG signal processing—for seizure detection \cite{ShKaMaJa,TrRuSa}, sensor optimization \cite{BaRuSaTr}, and dimensionality reduction \cite{AtQaGi}. In \cite{BeIaRo}, the High-Order SVD of tensors is used to estimate plant diversity with high accuracy and efficient memory use. 

The study in \cite{AdFaEl} introduces a denoising technique for biomedical signals that combines Independent Component Analysis (ICA) with a generalized Gamma distribution model. This hybrid method demonstrates robust performance in noise suppression for ECG and EEG signals, achieving improved signal clarity. A blind source separation algorithm for biomedical signals is proposed in \cite{ChZh}, using a gradient descent method on the Lie group manifold. By leveraging the geometric properties of the manifold, the method improves convergence speed, stability, and separation accuracy compared to classical ICA approaches.

Although Fourier Transform allows decomposition of signals into frequency components, it often lacks precision when filtering out specific wavenumbers. SVD, by contrast, allows separation of information via orthogonal subspaces, making it a valuable tool for extracting structural properties from signals or images \cite{MoStVa,WiDa}. From the SVD of a matrix $A$, the concept of Singular Entropy can be introduced to measure its information content. Prior work has shown that singular entropy is useful in areas such as noise reduction, vibration analysis, and image irregularity detection \cite{WiDa,YaTs}.

Recent studies continue to demonstrate the versatility and power of SVD and its variants in biomedical signal and image processing. For instance, \cite{ZhWa} explores the classification of ECG arrhythmias using signals reconstructed from lossy compression via Singular Value Decomposition (SVD). In \cite{104}, the authors present a technique combining Fast Independent Component Analysis (FastICA) with SVD to extract clean fetal ECG (FECG) signals from noisy abdominal recordings. A wavelet-based method is then used to detect QRS complexes and ST segments, achieving high levels of accuracy and signal clarity.

An extension of Singular Spectrum Analysis (SSA) is proposed in \cite{105}, where the method, named Multi-channel SSA-based Denoising (MSSAD), applies tensor decomposition to effectively denoise multi-channel time series. This approach leverages both intra-channel and inter-channel correlations to better separate noise from meaningful signals. SVD has also been applied to electromyography (EMG) processing. In \cite{106}, an SVD-based algorithm is introduced to remove cardiac interference from trunk EMG signals. The method operates in both time and frequency domains and does not require a reference ECG signal.

In the context of image processing, \cite{101} presents a method for image enlargement that combines SVD with cubic spline interpolation. The interpolation is applied to the SVD feature matrices, and the step size is optimized to maximize the signal-to-noise ratio (SNR). The work in \cite{102} proposes a modified SVD (MSVD) approach to enhance low-quality, low-resolution CCTV images. By introducing a threshold in the singular value matrix, the method improves image clarity and contrast.

The study in \cite{BeWiRu} proposes Reversed Auto-Encoders (RA), a generative approach that reconstructs pseudo-healthy counterparts of pathological images. This enables unsupervised anomaly detection across different medical imaging modalities, including brain MRI and pediatric X-rays, showing strong generalization capabilities. In \cite{Buttar}, a DCNN-based architecture is presented for detecting neurological anomalies in MRI scans, such as Alzheimer’s disease and epilepsy. The framework integrates preprocessing steps and cross-validation techniques, achieving high accuracy and robustness in medical anomaly detection.

While SVD is already a well-established tool in ECG analysis, particularly in the context of maternal-fetal signal separation, this paper proposes a new method based on Energy Gap Variation and Singular Energy to improve the selection of threshold values for more effective signal separation. In addition, GSVD is employed in this context, providing enhanced discriminative power and improved performance in separating fetal and maternal components.
Furthermore, we introduce a novel image analysis technique called Singular Smoothness, which combines Singular Entropy with the Frobenius norm to measure information density. This new framework is applied to detect and study natural anomalies such as mountain fractures and burned forest regions, enabling a comprehensive understanding of complex landscape patterns.

This paper is organized as follows. In Section \ref{sec2}, we present the mathematical preliminaries related to SVD and GSVD. Section \ref{sec3} focuses on medical signal analysis, where we apply Singular Entropy and the proposed Energy Gap Variation to improve maternal-fetal ECG separation. In this section, we also introduce the use of GSVD for even greater accuracy. Section \ref{sec4} develops the new concept of Singular Smoothness for image analysis, combined with the Frobenius norm to quantify information density. These tools are applied to real-world landscape images. Finally, Section \ref{sec5} presents numerical experiments that validate the effectiveness of our proposed methods.

\section{Preliminaries}\label{sec2}
According to the Theorem of SVD \cite{GoVa}, for a matrix $A \in {\mathbb R}^{m \times n}$,
there are two orthogonal matrices $U \in {\mathbb R}^{m \times m}$ and $V \in {\mathbb R}^{n \times n}$
and a diagonal matrix $\Sigma_r \in {\mathbb R}^{r \times r}$ such that
\begin{equation*} 
A = U \left[\begin{array}{cc} \Sigma_r & O \\O & O\end{array}\right]V^T,
\end{equation*}
where $r = {\rm rank}(A)$ and the bordered null-blocks may be absent. Moreover,
$\Sigma_r = {\rm diag}(\sigma_1,\dots,\sigma_r)$ with  $\sigma_1 \geq \dots \geq \sigma_r > 0$. In this case, the columns
$u_1,\dots,u_m$ of $U$ are called the left singular vectors while the columns $v_1,\dots,v_n$ of $V$ are the right
singular vectors. Following this notation, the matrix $A$ can be represented as
\begin{equation}\label{suma_SVD}
A = \sum_{i=1}^r \sigma_i u_i v_i^T,
\end{equation}
where $v_i^T$ stands for the transpose of the vector $v_i$.

We denote by $A_{LP}$ the low-pass subspace associated to the higher singular values
$\sigma_1, \dots, \sigma_m $, $A_{BP}$ the band-pass subspace associated to the residual $\sigma_{m+1}, \dots, \sigma_s$,
and $A_{HP}$ the high-pass subspace associated to the noise subspace. Under this notation,
the SVD allows us to separate the information projected in three different subspaces as follows:
\begin{equation}\label{separacion}
A = A_{LP} + A_{BP} + A_{HP}.
\end{equation}
The expression (\ref{separacion}) can be equivalently written as in (\ref{suma_SVD}) as follows
\[
A = \sum_{i=1}^m \sigma_i u_i v_i^T + \sum_{i=m+1}^s \sigma_i u_i v_i^T + \sum_{i=s+1}^r \sigma_i u_i v_i^T.
\]

SVD is a very powerful tool used for various applications such as dimensionality reduction, noise reduction, and least-squares fitting, etc. For instance, in cases where noise is spatially white and decorrelated from the signal, the subspace $A_{HP}$ can be identified as the noise subspace, and the remaining subspaces contain the most relevant information about the signal stored in $A$. Therefore, to reduce the noise, the matrix $A_{HP}$ should be deleted.

Another interesting application of SVD is the detection and comparison of landscape patterns. For instance, it is possible to study and quantify changes in vegetation patterns by analyzing aerial photographs \cite{Co}. Moreover, sometimes it is necessary to compare patterns to determine the texture of the landscape \cite{CoBaGa}. It is worth noting that in both cases, the starting point is an aerial image of the Earth's surface. Nowadays, images are typically stored in digital format, and representing them using matrices is a good approach. Then, Matrix Theory's methods can be applied to analyze the different characteristics of the information contained in an image. 

On the other hand, an alternative to the SVD is the GSVD which is applied when two data
matrices are given and allows us to compare information of each one
of them. According to the Theorem of GSVD \cite{GoVa}, for a matrix
$A \in {\mathbb R}^{m \times n}$ and a matrix $B \in {\mathbb R}^{s
\times n}$, with $m \geq n$, there are two orthogonal matrices $U
\in {\mathbb R}^{m \times m}$ and $V \in {\mathbb R}^{s \times s}$,
a nonsingular matrix $X \in {\mathbb R}^{n \times n}$, and diagonal
nonnegative matrices $C \in {\mathbb R}^{m \times n}$ and $S \in
{\mathbb R}^{s \times n}$ such that
\begin{equation*} 
A = U C X^T \qquad \text{ and } \qquad B = V S X^T,
\end{equation*}
where $C^T C+S^T S=I_n$.

We recall that the total energy in a $m$-vector's sequence {$a_k$},
 $k=1, \dots, n$, with associated $m \times n$ matrix $A$ of rank $r$ is
equal to the energy in the singular spectrum,

\[
\|A\|_F^2= {\rm trace}(A^TA) =\sum_{i=1}^m \sum_{j=1}^n a_{ij}^2.
\]
The oriented energy of $A$, measured in the direction $q$, is
defined as
\[
E_q[A]=
\| q^TA \|_F^2.
\]
In terms of elements, it can be computed as 
\[
E_q[A]  = {\rm trace} ( (q^TA)^T q^TA ) = {\rm trace}(\left[\begin{array}{ccc}
q^Ta_1 & \dots & q^Ta_n
\end{array}
\right] \left[\begin{array}{c}
q^Ta_1 \\ 
\vdots \\ 
q^Ta_n
\end{array}
\right]) = \sum_{k=1}^n (q^Ta_k)^2,
\]
where $a_k \in {\mathbb R}^m$ are the columns of matrix $A$, and $q \in
{\mathbb R}^{m \times 1}$ is a 2-unit vector. This is
a powerful tool to separate signals from different sources
\cite{MoStVa}. The columns $x_i$ of $X$, provided by the GSVD, are
vectors for which the oriented energy of matrix $A$ is
$\alpha_i$/$\beta_i$ times larger than the oriented energy of matrix
$B$, where $\alpha_i$ and $\beta_i$ are the singular values of $A$
and $B$ respectively. In \cite{CaMoVaSa} a separation method based
on the GSVD is proposed.

\section{ECG's analysis from a cutoff point}\label{sec3}
We focus on constructing subspaces that can separate the information of a known signal and display it distinctly. In this case, we use a vector to represent the information contained in the original signal. When a signal is given, it can be recorded in a vector as
\[
{\bf x} = \left[\begin{array}{cccc} x_1 & x_2 & \dots & x_m\end{array}\right].
\]
This vector can be reshaped into a matrix by selecting vectors of size $n$ and arranging them as columns of matrix $A$ as follows:
\[
A = \left[\begin{array}{ccccccc}
x_i & x_{j} & \dots & x_{k} \\
x_{i+1} & x_{j+1} & \dots & x_{k+1} \\
\vdots & \vdots & \ddots & \vdots \\
x_{i+n-1} & x_{j+n-1} & \dots & x_{k+n-1} \\
\end{array}\right] \in {\mathbb R}^{n \times p},
\]
where $n \leq m$ is the number of entries chosen in ${\bf x}$ and $p$ has to be (arbitrarily) selected.
It is well-known that
\[
A = U\Sigma V^T = \left[\begin{array}{ccc}  U_m & U_f & U_n \end{array}\right]
\left[\begin{array}{ccc}  \Sigma_m & O & O \\
                             O & \Sigma_f & O \\
                             O & O & \Sigma_n
                              \end{array}\right]
\left[\begin{array}{c}  V_m^T \\ V_f^T \\ V_n^T \end{array}\right],
\]
where $\Sigma_m$ contains $k_m$ singular values associated with the maternal heart,
$\Sigma_f$ contains $k_f$ singular values associated with the fetal heart, and
$\Sigma_n$ contains $k_n=p-k_m-k_f$ singular values associated with the noise and other possible sources of bioelectric
activity. Then, the ECG of the mother, fetus, and noise are respectively
\[
A_m =  U_m \Sigma_m V_m^T, \qquad A_f =  U_f \Sigma_f V_f^T, \qquad A_n =  U_n \Sigma_n V_n^T.
\]
Since the signal is not a perfect periodical signal, it does not
exist values of $k_m$ and $k_f$ that completely define the
subspaces. Then, the more important issue is to find the most optimum values that make this job.

\subsection *{Energy gap variation}

Let $A$ be a matrix containing two signals, one corresponding to the maternal heart and the other corresponding to the fetal heart. As the energy of the maternal signal is much stronger than that of the fetal signal, the aim is to separate both signals in matrix $A$ by analyzing the energy distribution. This study can be extended to a matrix containing three signals (maternal, fetal, and noise), but, for the sole purpose of illustrating the technique, we first consider only the first two signals. Alternatively, we could consider both fetal and noise signals together to separate the maternal signal, and then apply the same method to separate the noise from the fetal signal. To model the effect of noise in biomedical signals such as ECG, we assume that the additive noise is independent and identically distributed (i.i.d.) Gaussian with zero mean and constant variance, and uncorrelated with the signal of interest. Under this statistical model, the smallest singular values in the decomposition are interpreted as primarily representing noise components. This provides a theoretical foundation for the use of SVD and GSVD in noisy signal environments. Moreover, it enables a principled approach to thresholding the singular value spectrum, which is essential for accurate signal-noise separation. In this context, the proposed Energy Gap Variation (EGV) and Singular Energy metrics offer a probabilistic framework for identifying the cutoff point between dominant signal components and background noise. 
Under these conditions, matrix $A$ can be expressed as

\[
A = U\Sigma V^T = \left[\begin{array}{cc}  U_m & U_f
\end{array}\right]
\left[\begin{array}{cc}  \Sigma_m & O \\
                             O & \Sigma_f \\
                            \end{array}\right]
\left[\begin{array}{c}  V_m^T \\ V_f^T  \end{array}\right].
\]

Let $M$ and $F$ be the matrices corresponding to heart signals, that
can be reconstructed by means of

\[
M = \sum_{i=1}^m \sigma_i u_i v_i^T,
\quad \quad
F = \sum_{i=m+1}^r \sigma_i u_i v_i^T,
\]
where $r$ is the rank of $A$.

Denote
\[
M_k = \sum_{i=1}^k \sigma_i u_i v_i^T, \quad \text{ for } 1 \leq k
\leq m
\]
and
\[
F_p = \sum_{i=p}^r \sigma_i u_i v_i^T, \quad \text{ for } m+1 \leq p
\leq r.
\]

To separate the maternal and fetal signals in the ECG, we need to determine the value of $m$. We can find this value by analyzing the energy of matrix $M_k$ and identifying the corresponding value of $k$.

Since we know that the energy of $M$ is much bigger than the energy
of $F$, the procedure consists of comparing the energy of $M_k$ and
$F_{k+1}$ by a sliding window method. For different values of $k$,
we will get different energy values for $M_k$ and $F_{k+1}$.

The energy of $A$ can be expressed as
\[
E_A = \|A\|_F^2 = \|M_k\|_F^2 + \|F_{k+1}\|_F^2
\]
where
\[
\|M_k\|_F^2=\left\|\sum_{i=1}^k \sigma_i u_i
v_i^T\right\|_F^2=\sum_{i=1}^k \sigma_i^2
\]
\[
\|F_{k+1}\|_F^2=\left\|\sum_{i=k+1}^r \sigma_i u_i
v_i^T\right\|_F^2=\sum_{i=k+1}^r\sigma_i^2.
\]

For each value of $k$, the energy gap (difference of energy) between
$M_k$ and $F_{k+1}$ is given by
\[
G_k = \|M_{k}\|_F^2 - \|F_{k+1}\|_F^2
\]

The increasing of energy gap is
\begin{eqnarray*}
\Delta G_k & = & (\|M_{k+1}\|_F^2 - \|F_{k+2}\|_F^2) - (\|M_k\|_F^2 - \|F_{k+1}\|_F^2) \\
              & = & \left(\sum_{i=1}^{k+1} \sigma_i^2 - \sum_{i=k+2}^r \sigma_i^2\right) -
              \left(\sum_{i=1}^k \sigma_i^2  -  \sum_{i=k+1}^r \sigma_i^2\right) \\
              & = & 2\sigma_{k+1}^2.
\end{eqnarray*}

Finally, to calculate the EGV we propose to use the concept of
singular energy, based on the singular entropy \cite{YaTs}. The
singular energy is given by

\[
SE_i = - \frac{\Delta G_i}{\gamma} \ln\left(\frac{\Delta
G_i}{\gamma}\right) \qquad \text{ with } \quad \gamma = \sum_{j=1}^r
\Delta G_j
\]
for each $i=1,\dots,r$, and we define
\[
V(\Delta G_k)= SE_k - SE_{k-1}.
\]



When the singular values corresponding to the maternal signal are in
$M_k$ and the singular values corresponding to the fetal signal are
in $F_{k+1}$, the biggest EGV will occur. In this situation, the
maximum difference between two consecutive singular values occurs
between the smallest maternal singular value and the biggest fetal
singular value, thus, causing a maximum in the EGV.
When this happens, $m=k$ and the maternal signal can be
reconstructed from the first $m$ singular values of $A$.

Then,
\[
\max_{1 \leq k \leq r-1} V(\Delta G_k) = V(\Delta G_m) = SE_m -
SE_{m-1}.
\]

This method can be simplified due to the symmetry of the problem.
Denote

\[
A_k = \sum_{i=1}^k \sigma_i u_i v_i^T.
\]

The increasing of the energy gap between $A_k$ and $A_{k+1}$ can be
expressed as

\[
\Delta GA_k = \|A_{k+1}\|_F^2 - \|A_k\|_F^2 = \sigma_{k+1}^2 =
\frac{1}{2}\Delta G_k,
\]

and the EGV is given by



\[
V(\Delta GA_k) = \frac{1}{2}V(\Delta G_k) = \frac{1}{2}SE_k -
\frac{1}{2}SE_{k-1}.
\]

Then,

\[
\max_{1 \leq k \leq r-1} V(\Delta GA_k) = V(\Delta G A_m) =
\frac{1}{2} V(\Delta G_m) = \frac{1}{2}\left( SE_m -
SE_{m-1}\right).
\]

Suppose now that matrix $A$ also contains noise. If the ECG signal is recorded properly, the energy of the noise signal should be much smaller than that of the fetal signal. This implies that the singular values corresponding to the noise signal are those with the smallest values.

To perform spectral separation of maternal, fetal, and noise signals, we need to identify the values of $m$ and $f$ by finding two peaks in $V(\Delta GA_k)$. These peaks will correspond to the separation of the signals in the frequency domain. The first peak will occur between the maternal and fetal signals, and the second peak will occur between the fetal and noise signals. The value of $k$ at which the first peak occurs will correspond to $m$, while the value of $k$ at which the second peak occurs will correspond to $f$.

\begin{quote}
{\bf\sc Algorithm 1}

{\sl Inputs}: Matrix $A \in {\mathbb R}^{n \times n}$.

{\sl Outputs}: Cutoff point $m$.

\begin{description}
\item[Step 1] \; Compute SVD($A$).

\item[Step 2] \; Calculate $r$ (from size of matrix $S$).

\item[Step 3] \; Compute the singular values of $A$.

\item[Step 4] \; Compute $V(\Delta G_k)$ for $k=1,\dots,r-1$, using
the singular values of $A$.

\item[Step 5] \; Find the maximum $V(\Delta G_k)$.

\item[Step 6] \; Set $m=k$ for $V(\Delta G_k)$ of previous step.

\item[End]
\end{description}
\end{quote}

Another alternative technique that can be applied is based on the GSVD. In the method proposed in \cite{CaMoVaSa}, matrix $A$ contains maternal and fetal signals, while matrix $B$ contains only maternal QRS-intervals that do not coincide with fetal complexes. This method involves the projection of recorded signals (matrix $A$) onto some directions $x_i$, resulting in maternal ECG-free fetal signals. We propose a new method named generalized energy gap variation (GEGV), based on the EGV. The difference between both methods lies in the singular values concept used. While the EGV uses the classical singular values, for GEGV we use the generalized singular values for matrix $A$ given by the GSVD. Note that this method also works when matrix $B$ contains both maternal and fetal signals. However, since matrix $A$ contains a stronger fetal signal (electrodes placed on the abdomen) than matrix $B$ (electrodes placed on the thorax), we obtain better results reconstructing the fetal signal from matrix $A$.

It is important to remember that the position of the electrodes and the quality of the recorded signal are crucial factors to consider. Therefore, the cutoff point used to reconstruct the fetal signal from matrix $A$ is determined by


\[
\max_{1 \leq k \leq r-1} V(\Delta GA_k) = V(\Delta G A_m) =
 SE_m - SE_{m-1},
\]

where

\[
A_k = \sum_{i=1}^k \sigma_i u_i v_i^T,
\]

and $\sigma_i$, for $i=1,\dots,r$ are the generalized singular
values of matrix $A$, and the rank of matrix $A$ is $r$.

\begin{quote}
{\bf\sc Algorithm 2}

{\sl Inputs}: Matrices $A$ and $B \in {\mathbb R}^{n \times n}$

{\sl Outputs}: Cutoff point $m$.

\begin{description}
\item[Step 1] \; Compute GSVD($A$,$B$).

\item[Step 2] \; Calculate $r$ (from size of matrix $C$ from GSVD($A$,$B$).

\item[Step 3] \; Compute the generalized singular values of $A$.

\item[Step 4] \; Compute $V(\Delta G_k)$ for $k=1,\dots,r-1$, using
the generalized singular values of $A$.

\item[Step 5] \; Find the maximum $V(\Delta G_k)$.

\item[Step 6] \; Set $m=k$ for $V(\Delta G_k)$ of previous step.

\item[End]
\end{description}
\end{quote}

Both situations, based on SVD and GSVD, will be illustrated with
numerical experiments in Section \ref{sec5}.

\section{Landscape study}\label{sec4}
In this section, we apply the SVD to an image represented by a matrix $A$. In the context of image analysis, we assume that the observed data matrix can be modeled as the sum of a low-rank structured signal and additive noise. The noise is also modeled as an independent and identically distributed (i.i.d.) Gaussian process with zero mean and constant variance. Under this assumption, the smallest singular values of the image matrix are primarily associated with noise or fine-scale texture, while the largest singular values correspond to dominant structural features. This statistical model supports the interpretation of the singular value spectrum as a tool for anomaly detection. In particular, the proposed Singular Entropy and Singular Smoothness metrics allow us to quantify irregularities in the distribution of singular values, providing a sensitive indicator of disrupted or anomalous regions, such as mountain fractures or burned forest areas, based on deviations from the expected statistical behavior of smooth natural landscapes.
We recall that matrix $A$ can be decomposed as in (\ref{suma_SVD}), that is, the image can be obtained from the sum of subimages of the same size.
Each one of these subimages $u_iv_i^T$ is pondered by the singular value $\sigma_i$.

On the other hand, texture characterization has been studied from different points of view \cite{Co,CoBaGa}. For example, in \cite{WiDa}, a tool called information density is defined based on the Frobenius norm as follows:

\begin{equation}\label{id}
I(D) = \sqrt{\sum_{i=min}^{max} \sigma_i^2},
\end{equation}
where $min$ and $max$ are values to be determined and $D$ is a given
image. This concept is combined with the method of moving windows in
order to obtain the rugosity of the landscape. This technique is
effective when the whole image presents uniform brightness. This is
due to the fact that information density only considers the
magnitude of the selected singular values instead of their relative
importance.

In order to achieve greater independence of the result with respect to image brightness, this section introduces the concept of singular smoothness. For a given fragment $D$ extracted from the overlapping region of a window and the image, the $n$th order singular smoothness is defined as:
\begin{equation}\label{ss}
S_n(D) = \sqrt{\sum_{i=1}^{n}
\frac{\sigma_i^2-\sigma_{i+1}^2}{\sigma_{i+1}^2}}
\end{equation}


where $\sigma_1$ represents the largest singular value in the SVD of $D$. The purpose is to provide information that relates the values of different singular values in order to achieve brightness independence, and the number of computations required depends on the energy of each singular value.

In general, using singular smoothness of order 1 yields good results. However, for improvement, we can select a suitable order-n for the singular smoothness by analyzing the energy of the singular values. We can determine the appropriate order by identifying when two consecutive singular values are very close, i.e., $\sigma_i \approx \sigma_{i+1}$. The value of $i$ will indicate the order-n.

\begin{quote}
{\bf\sc Algorithm 3}

{\sl Inputs}: Matrix $D \in {\mathbb R}^{m \times n}$, $\delta$

{\sl Outputs}: $n$, $S_n(D)$.

\begin{description}
\item[Step 1] \; Compute SVD($D$).

\item[Step 2] \; Calculate $r$ (from size of matrix $S$).

\item[Step 3] \; Compute the generalized singular values of $D$.

\item[Step 4] \; Find $n$ such that $\sigma_n - \sigma_{n+1} \leq
\delta$.

\item[Step 5] \; Compute $S_n(D$).

\item[End]
\end{description}
\end{quote}

 Numerical experiments are presented in
next section. 

\section{Results of the numerical experiments}\label{sec5}

In this section we present numerical experiment for ECG signal
analysis and image analysis. We have used the MATLAB R2024 package.

\subsection{ECG}\label{ecglabel}
Electrodes placed on the body surface of a pregnant woman can capture signals with electrical activity of the fetal and maternal hearts. By recording thorax and abdominal signals, we can construct the matrix $A$ for SVD or matrices $A$ and $B$ for GSVD.

\subsection*{Matrix $A$ contains abdominal and thorax information}

\begin{figure}[h]
\centering
\includegraphics[height=4.5cm]{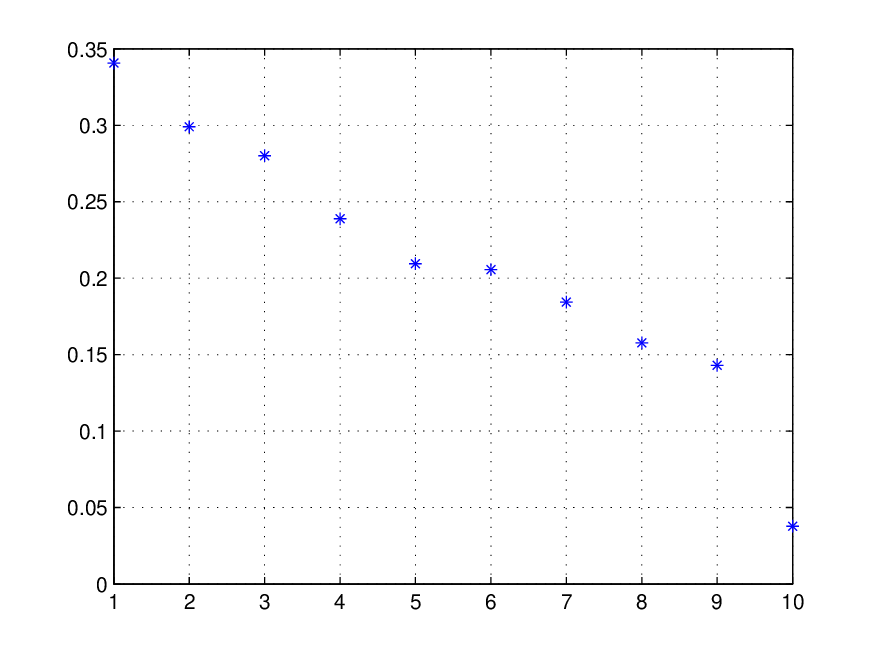}
\includegraphics[height=4.5cm]{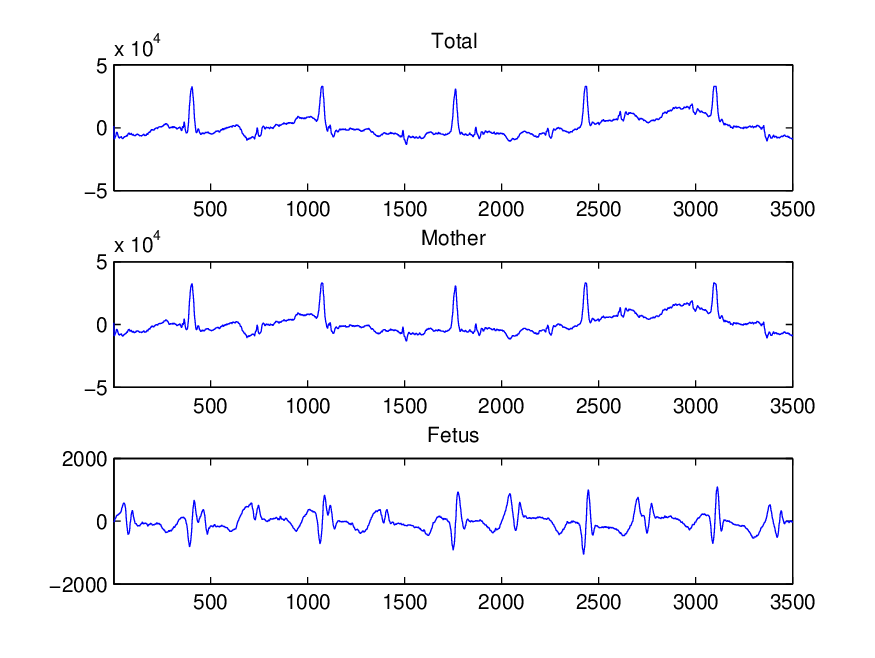}
\caption{Energy gap variation and ECG-MECG-FECG.}\label{se}
\end{figure}

The matrix $A$ has been constructed from 4 abdominal signals and 2
thorax signals. Each one of the 10 columns of $A$ consists of 3500
samples of the preceding signals. Figure \ref{se} displays the EGV
corresponding to matrix $A$. Applying the Algorithm 1 to get a
cutoff point, we determine that $k_m=4$ and $k_f=9$.
Figure \ref{se} also displays the total
signal (abdominal plus thorax signals) and the reconstructed MECG
and FECG applying the obtained $k_m$ and $k_f$ limit values. It can be seen
that the fetal heart rate is approximately double than the maternal
one.

\subsection*{Matrix $A$ contains thorax information and matrix $B$ abdominal information}

The matrix $A$ was constructed from 3 abdominal signals, and matrix $B$ was constructed from 3 thorax signals. Each of the 10 columns in $A$ and $B$ consists of 3500 samples of the respective signals.

\begin{figure}[h]
\centering

\includegraphics[height=4.5cm]{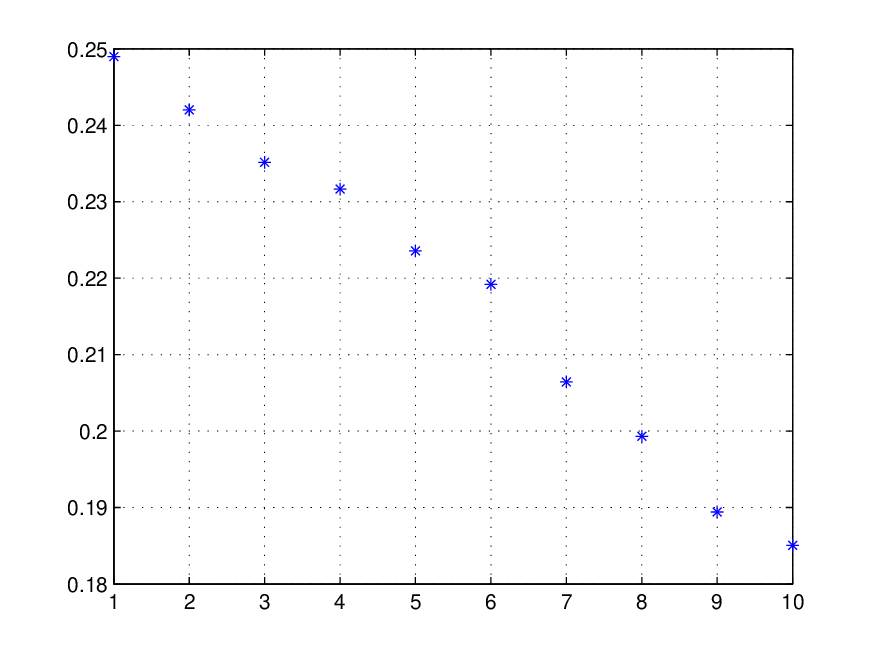}
\includegraphics[height=4.5cm]{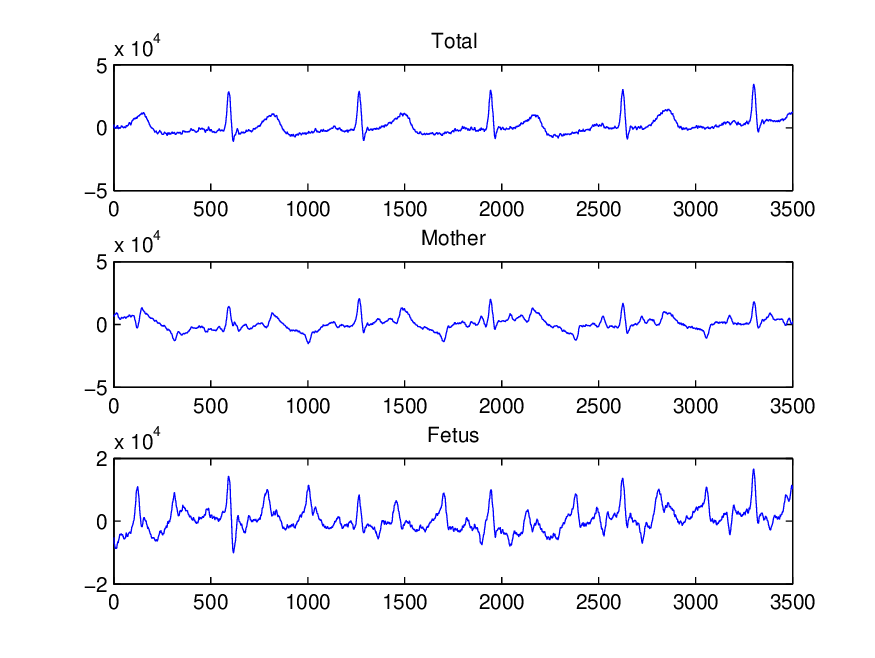}
\caption{GSVD energy gap variation and ECG-MECG-FECG.}\label{gsvdse}
\end{figure}

To separate the MECG and FECG signals, we obtained the singular values of matrices $A$ and $B$ using Algorithm 2. As shown in Figure \ref{gsvdse}, the separation of the signals ($k_m=6$) is determined by the intersection of the singular entropies of $A$ and $B$. This figure also displays the total signal and the reconstructed MECG and FECG. In this new situation, it can be observed that the fetal heart rate is also higher than the maternal one.

\begin{figure}[ht]
\centering
\includegraphics[height=4.5cm]{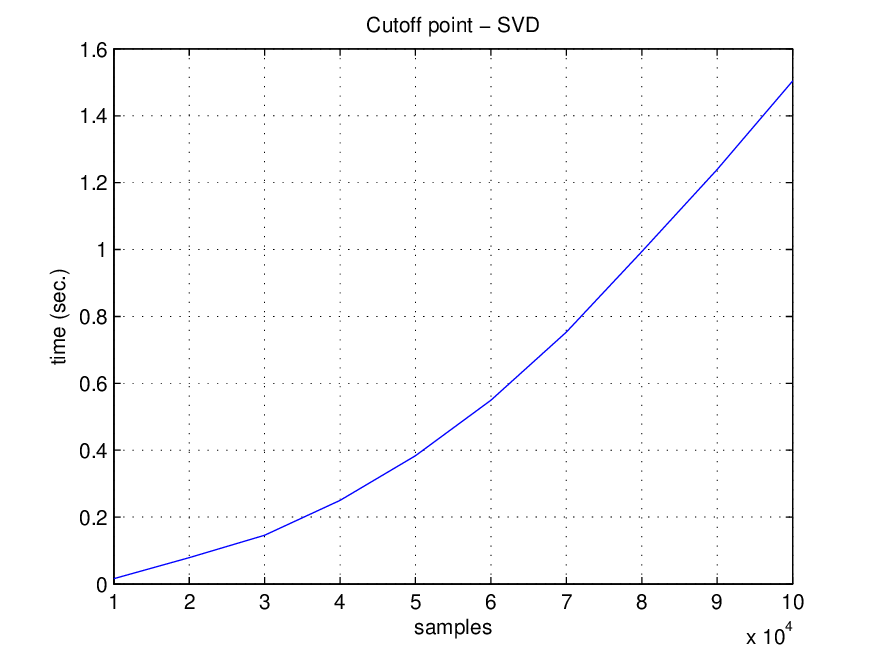}
\includegraphics[height=4.5cm]{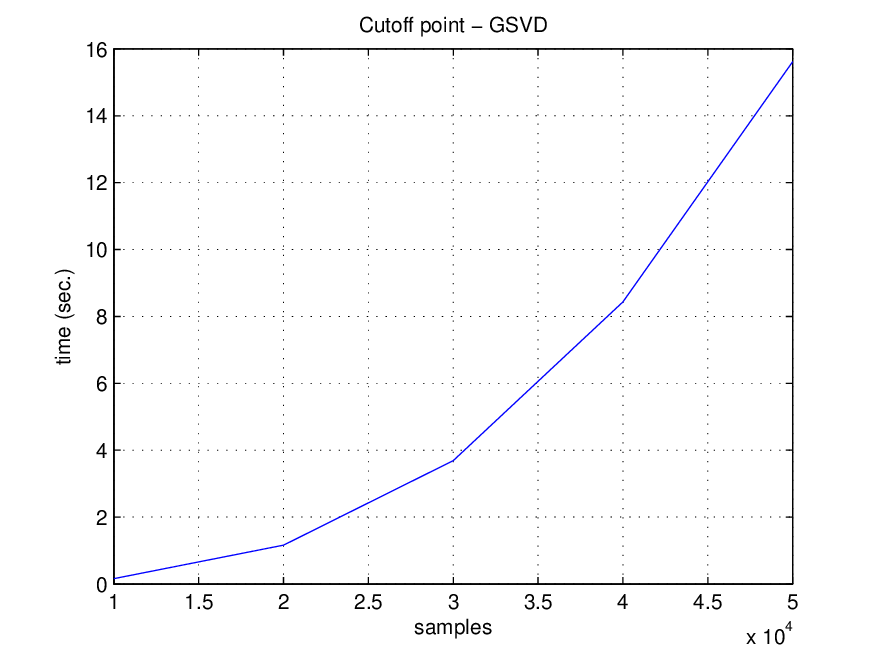}
\caption{Cutoff point computation time: SVD and GSVD methods.}\label{svdtime}
\end{figure}

Figure \ref{svdtime} shows the computational times for both cutoff methods, based on SVD and GSVD. It is clear that the time increases with the number of samples recorded in matrices $A$ and $B$. As expected, in the case of the GSVD method, the time increases much more than with the SVD method.


\subsection{Landscape}\label{landlabel}

In order to get results independently from the luminosity,
our method (\ref{ss}) uses the concept of singular smoothness instead of
density of information. If the image is plane, $\sigma_1$ has a high
value while the other singular values are almost zero. On the other
hand, in case of a random image, the information is distributed
among all the singular values. Expression (\ref{id}) defining the information density does not take into
account the information distribution, but only the amount of
information in some singular values. However, with
singular smoothness, for example of order 1, we get the amount of
information of $\sigma_1$ with respect to $\sigma_2$. In this way,
the singular smoothness indicates the degree of smoothness in the
image.


\begin{figure}[h]
    \centering

    \begin{subfigure}[t]{0.45\textwidth}
        \centering
        \includegraphics[height=4.5cm]{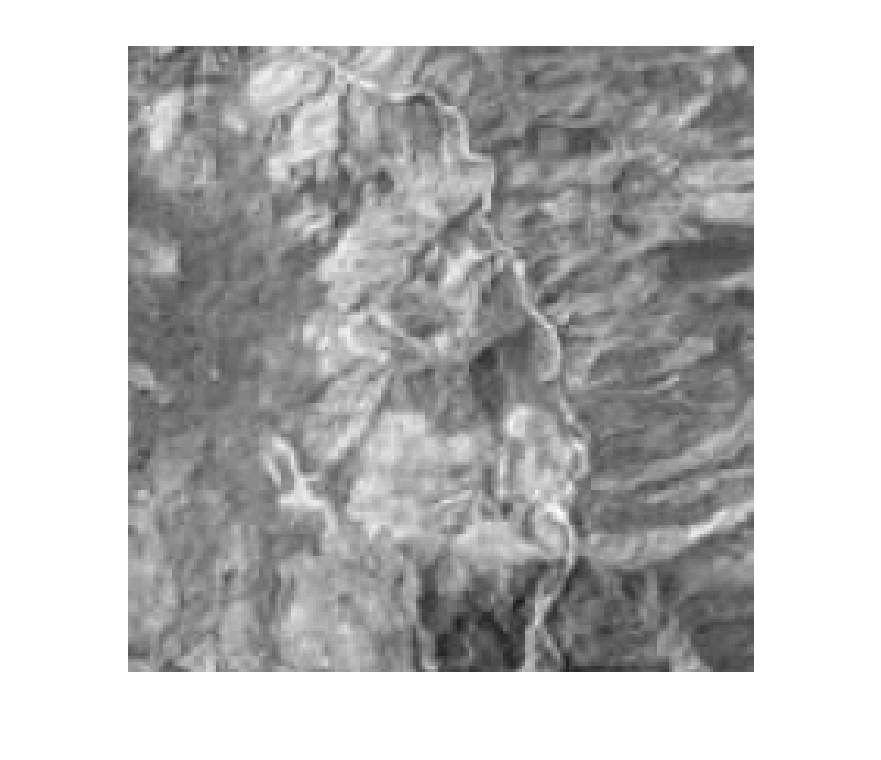}
        \caption{} 
        \label{fig:landscape}
    \end{subfigure}%
    \begin{subfigure}[t]{0.45\textwidth}
        \centering
        \includegraphics[height=4.5cm]{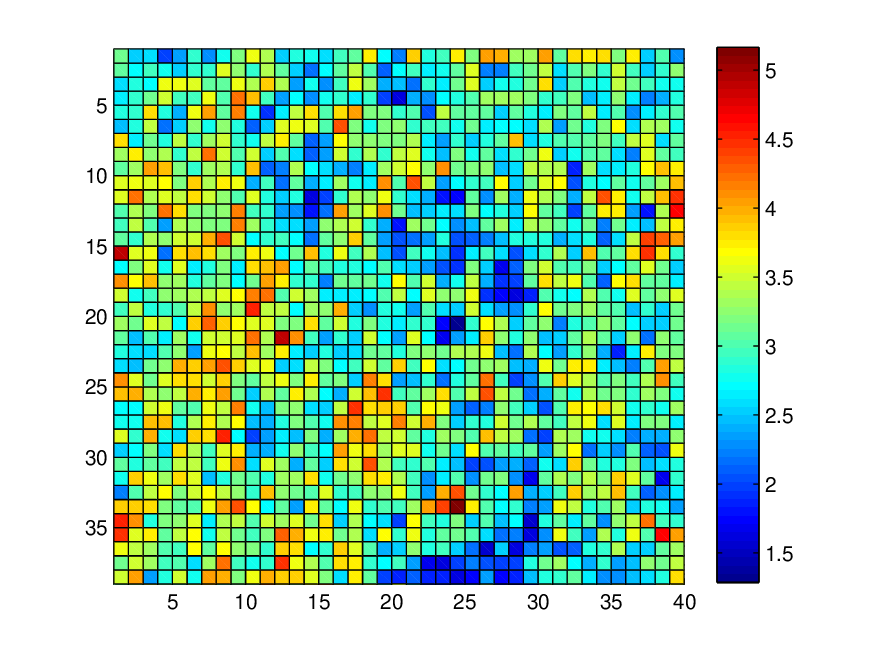}
        \caption{} 
        \label{fig:crack}
    \end{subfigure}

    \caption{Forest and burn severity. (a) Original forest. (b) Burn severity map.}
    \label{gs}
\end{figure}

\begin{figure}[h]
    \centering

    \begin{subfigure}[t]{0.45\textwidth}
        \centering
        \includegraphics[height=4.5cm]{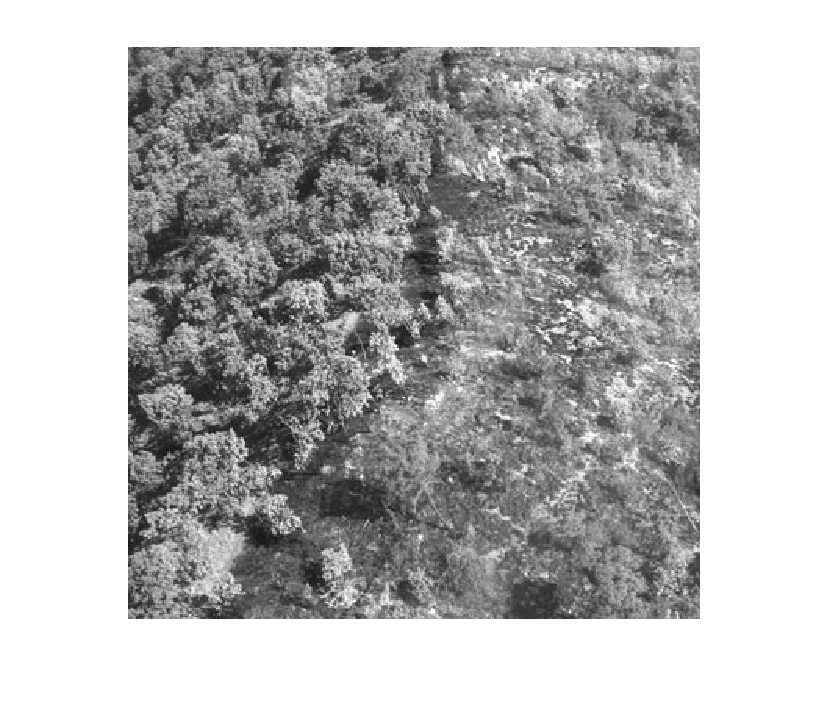}
        \caption{} 
        \label{fig:landscape}
    \end{subfigure}%
    \begin{subfigure}[t]{0.45\textwidth}
        \centering
        \includegraphics[height=4.5cm]{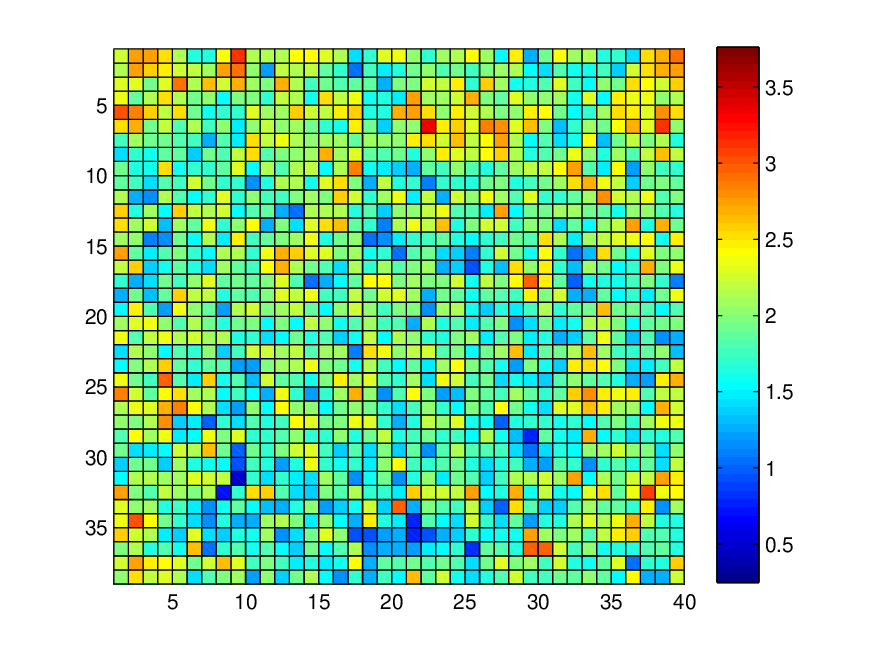}
        \caption{} 
        \label{fig:crack}
    \end{subfigure}

    \caption{Forest and burn severity. (a) Original forest. (b) Burn severity map.}
    \label{bs}
\end{figure}

Using Algorithm 3, we can identify cracks in a landscape, as shown in Figure \ref{gs}. By applying the singular smoothness method with order 2, the cracks can be highlighted in blue. One advantage of this method over the information density method is the reduced number of operations required. In this example, we have performed 2500 operations compared to 39920 for the density method. Another example is shown in Figure \ref{bs}, where a partially burnt forest is displayed. It can be observed that the fragments representing burnt areas have a high value compared to those from unburnt areas, due to the increased variation caused by foliage. Some isolated blue squares representing foliage can also be seen within the red squares representing burnt zones. This is because of the presence of stones in the image. In positions where the sliding window includes both burnt areas and stones, the variation is high, resulting in low values of smoothness.

\begin{figure}[h]
\centering
\includegraphics[height=4.5cm]{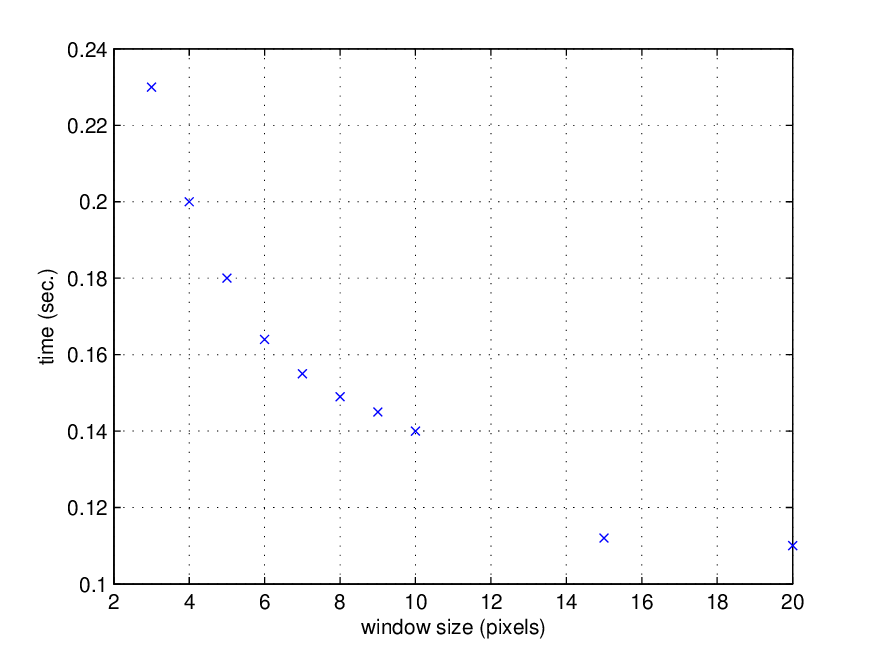}
\caption{Singular smoothness computation time.}\label{sstime}
\end{figure}

In Figure \ref{sstime}, computational time for image analysis based
on singular smoothness is displayed. Selecting bigger windows will
decrease the computational time, because of less SVD computations
are performed, but will provide less resolution in the results. In
the examples shown in Figures \ref{gs} and \ref{bs}, the size of the
windows is 5 $\times$ 5 pixels.


\section{Conclusions}\label{sec6}

We have introduced new methods for signal and image analysis based on the SVD and GSVD decompositions. We applied singular entropy to signal analysis to obtain a new cutoff point for the singular values. We also introduced the concept of singular smoothness, which takes into account the amount of information present in the signal or image. The statistical modeling of additive noise as an i.i.d. Gaussian process has guided our interpretation of the singular value spectrum in both ECG signal and image analysis. This framework justifies the use of SVD and GSVD for separating signal from noise and underpins the design of our proposed metrics: EGV and Singular Energy for biomedical signals, and Singular Entropy and Singular Smoothness for image-based anomaly detection. This approach differs from the classic concept of density information, which only considers the amount of information in some singular values and does not take into account its distribution. This novel approach has allowed us to perform image analysis more effectively. Moreover, although the present work has focused on biomedical signals and landscape images, the proposed algebraic and statistical framework can be naturally extended to other application domains. Problems involving the separation of signal from noise or the detection of structural irregularities, such as in audio processing, seismic exploration, financial time-series, or remote sensing, could also benefit from the introduced concepts of Energy Gap Variation, Singular Entropy, and Singular Smoothness.


%
%


\end{document}